\documentclass[useAMS,usenatbib]{mn2e}

\usepackage{graphicx}

\title[Formation of interstellar propylene]
{A radical route to interstellar propylene formation}

\author[J.M.C. Rawlings et al.]
{J.M.C.~Rawlings,$^1$\thanks{E--mail: jcr@star.ucl.ac.uk}
D.A.~Williams, $^1$
S.~Viti,$^1$
and  C.~Cecchi--Pestellini$^2$\\
$^1$University College London, Department of Physics and Astronomy,  
Gower Street, London WC1E 6BT, United Kingdom\\
$^2$INAF -- Osservatorio Astronomico di Palermo, P.za
Parlamento 1, I-90134 Palermo, Italy}

\begin{document}

\maketitle

\begin{abstract}
Complex organic molecules, such as propylene (CH$_3$CHCH$_2$), are detected in
molecular clouds (such as TMC1) with
high fractional abundances ($\sim 2\times 10^{-9}$, relative to hydrogen) that
cannot be explained by gas-phase chemical reactions under normal dark cloud
conditions.
To obtain such high abundances requires an efficient grain-driven chemistry to
be operating, coupled with effective desorption of the complex organics back 
into the gas-phase.
We propose that the mechanism that we have previously described; rapid
high density gas-phase chemistry in the gas released following sudden, total, 
ice mantle sublimation - can explain the high abundances, without recourse to 
ill-defined surface chemical pathways.
Our model provides a natural explanation for why it is that some sources 
harbour high abundances of propylene, whilst others do not; based on the 
age and level of dynamical activity within the source (which affects the ice
composition) and the chemical composition of the ambient gas.
\end{abstract}


\section{Introduction}

Propylene, CH$_3$CHCH$_2$, also known as propene, has been detected in dense 
gas towards the cyanopolyyne peak of the low mass 
star-forming region TMC-1 with a substantial fractional 
abundance of $\sim 2\times 10^{-9}$ relative to hydrogen \citep{M07}. However, 
the same paper also states that searches for propylene towards the massive 
star-forming region Orion KL have been unsuccessful. 
Detailed modelling studies \citep{HRT10,Lin13} suggest that propylene cannot 
be readily formed by conventional interstellar gas phase chemistry. 
Therefore, the relatively high abundance of propylene in TMC-1 appears to 
be unexplained. This letter considers a possible gas-phase chemical 
route to the 
formation of propylene as an alternative to conventional interstellar 
chemistry. This new route may also contribute a variety of complex organic 
molecules (COMs) to interstellar clouds. Of course, this possible 
alternative route does not replace conventional interstellar chemistry, 
but it may supplement it in certain situations.
While others have also invoked grain surface radical-radical reactions 
in certain situations \citep[e.g.,][]{GWH08} it appears to be difficult
to establish high abundances of product molecules in normal (dark cloud)
interstellar conditions.

\section{Radical Associations}

The structure of propylene suggests that the molecule may be formed by the
addition of radicals CH$_3$, CH$_2$, and CH in a suitable network of reactions, 
such as
\[
{\rm CH_3 + CH  \to CH_3CH}
\]
followed by
\[
{\rm CH_3CH + CH_2 \to CH_3CHCH_2.}
\]
\citet{W71} first discussed radical reactions in the context of 
interstellar chemistry. Subsequently, \citet{MW75} showed that radicals 
of more than a few atoms may associate at near collisional rates. The 
resulting association complexes stabilise radiatively or, if the density is
sufficiently high, in 3-body collisions. In their feasibility study they found
that a chemistry based solely on radiative association of radicals 
has the potential to develop a significant population of molecules 
containing up to about ten atoms. It is now accepted that radiative 
associations of radicals play an important part in all models of 
conventional gas phase interstellar chemistry. However, it is apparent 
that these conventional schemes do not lead to the production of 
substantial amounts of propylene.

We have recently explored a radical addition chemistry that may occur in
very high density gas generated by the explosive evaporation of interstellar
ice mantles on dust grains \citep{RWV13}. These explosions are
widely observed in the laboratory, and are believed to be generated in the
interstellar medium (ISM) by the accumulation of H-atoms in the ices until a
critical H-atom density is reached, when a runaway explosion occurs 
\citep{DW11}. Such explosions were first noted in the astrochemical
context by \citet{G76} in his laboratory experiments. \citet{RWV13}
assume that radicals such as OH, CH$_3$, HCO, etc., are created in the
ice mantles during their build-up on grains by the action of the cosmic ray
induced radiation field on molecules in the ice \citep{PT83}.
\citet{RWV13} showed that an explosion chemistry of
radical associations stabilised in 3-body reactions may  persist for
timescales on the order of nanoseconds in the expanding gas, and that this
chemistry may produce significant amounts of COMs. 

However, Rawlings et al. did not examine the variety of chemical species 
likely to be produced in such a scheme. In this Letter, we present and discuss
the predictions for the rich post-explosion chemistry. 
This chemistry is assumed to be entirely due to radical associations 
stabilised by 3-body collisions in which H$_2$O is the dominant stabilising 
partner. 

\section{Physical and chemical model}

\subsection{Physical model of the explosion}

The physical basis for our model is the same as that described in \citet{RWV13}; 
we assume that the chemical processes occur in the gas-phase, in the high density 
gas that is produced as a result of the sudden, and total, sublimation of ice
mantles. In this particular study we do not speculate as to the cause of this 
mechanism, we simply note that any sudden, efficient, and repetitive desorption
process is capable of driving the sort of chemistry that we describe above.

As in \citet{RWV13} we consider an idealised situation in which a sphere of
ice is instantaneously sublimated into the gas-phase, and then subsequently 
undergoes free expansion into a vacuum at some fraction, $\epsilon$, of the 
sound speed $v_s$. 
The parameter $\epsilon$ compensates for real grain morphologies and makes some 
allowance for deviations from spherical symmetry and/or the effects of trapping 
in cavities. Unhindered spherical expansion corresponds to $\epsilon=1$.

If the gas sphere has initial radius $r_0$ and density $n_0$ (which may be 
comparable to the solid-state number density) then, by mass conservation, at any 
time $t$ after mantle sublimation, the density $n$ is given by
\begin{equation}
\frac{n}{n_0} = \frac{1}{\left( 1+10^{9}\epsilon t \right)^3}
\label{eqn:dynamics}
\end{equation}
where we have assumed that $r_0$ is comparable to the typical thickness of an 
ice mantle ($r_0 = 10^{-5}$~cm), and the local sound speed
$v_s=10^4$~cm\,s$^{-1}$.

Table~\ref{tab:parameters} summarises the range of values assigned to the 
various physical parameters in the model.

\begin{table}
\begin{center}
\caption{Physical and chemical parameters defining the explosion. See text for 
description.}
\label{tab:parameters}
\begin{tabular}{lc}
\hline
Parameter  & Value \\ 
\hline
CH$_4$/H$_2$O & 0.04 \\
NH$_3$/H$_2$O & 0.01 \\
H$_2$CO/H$_2$O & 0.03 \\
CH$_3$OH/H$_2$O & 0.03 \\
Fraction converted into radicals ($F_{rad}$) & 0.01-1\% \\
\hline
Initial density ($n_0$) & $10^{19}-10^{23}$ cm$^{-3}$ \\
Final density $(n_{fin}$) & $n_0$/1000 \\
Three-body rate coefficients ($k_{3B}$) & $10^{-32}-10^{-26}$ cm$^6$s$^{-1}$ \\
Trapping parameter ($\epsilon$) & 0.1-1.0 \\
\hline\hline
Molecule $\to$ radicals & branching ratio \\ 
\hline
H$_2$O $\to$ OH & 0.5 \\
CH$_4$ $\to$ CH$_3$, CH$_2$, CH & 0.33,0.33,0.33 \\
NH$_3$ $\to$  NH$_2$, NH & 0.5,0.5 \\
H$_2$CO $\to$ CHO & 0.5 \\
CH$_3$OH $\to$ CH$_3$O, OH, CH$_3$, CH$_2$OH & 0.1,0.2,0.2,0.5 \\
\hline
\end{tabular}
\end{center}
\end{table}

\subsection{Chemistry produced from exploding ices}

We shall assume for simplicity that the ice consists mainly of H$_2$O, CO, 
CO$_2$, CH$_4$, NH$_3$, H$_2$CO, and CH$_3$OH; we recognise that other 
species (in particular, CN bearing species) are also likely to be present 
in the ice, but at lower abundance levels. 
The relative abundances of these species are known to vary from one line 
of sight to another \citep[e.g.,][]{Getal04,Betal04}. We adopt as canonical 
values those shown in Table~\ref{tab:parameters};
we also consider variations from those values. The ice 
mantles suffer photodissociations driven by the cosmic ray induced 
radiation field.
\citet{RWV13} showed that an upper limit (before mantle explosion occurs) for
the fraction of mantle species that has been photodissociated is of the order 
of 1\%. In our models we have therefore considered values in the range 0.01-1\%.
The assumed products of the photodissociations and their 
relative abundances are also shown in Table~\ref{tab:parameters}.
These products are assumed to be 
retained in the ices until explosions occur. Once the ice mantle on a 
particular grain has exploded, we assume that it begins to accumulate 
another ice mantle, and the process repeats.
Note that this list only includes reactive radicals. Thus, although they 
constitute significant components of interstellar ices, we have not included 
CO or CO$_2$ as these species will not be dissociated into chemically active
reactants.

Table~\ref{tab:parameters}
shows that we omit from the photodissociation products the atoms
such as O, C, and N on the grounds that the overabundance of hydrogen will
tend to enhance the hydrogenation. Hot hydrogen atoms in the explosion 
will tend to establish a population of hydrides.

The first stage of the radical addition chemistry considered here involves
association between the nine radical species identified in 
Table~\ref{tab:parameters}.
Note that $\sim$99\% of the molecules from the ices are not involved, 
except to provide third bodies stabilising the products. 
The first stage generates a set of 45 associations, giving products that are
either molecules (26 of these, formed when the initial radicals have equal
valences) or radicals (of which there are 19, formed where the initial radicals
have unequal valences).
The product molecules are generally familiar and many of them are detected
interstellar species. It is assumed that these product molecules take no 
further part in the radical chemistry. However, product radicals from the first
stage may undergo further stages of association. The product radicals of the
first stage capable of forming propylene in the second stage are CH$_3$CH and 
CH$_2$CH. 

Just as in the first stage, associations in the second stage may give rise 
either to saturated molecules or to new radicals. The new radicals may, if 
the expansion timescale permits, give rise to a further stage of 
chemistry. We do not study this additional stage in our model.
The 19 radicals generated in the first stage may associate with the 9 initial
radicals to form 94 saturated molecules. 
An example of the reactions in the two stages is given in 
Table~\ref{tab:reactions}. A full listing can be provided on request.

\begin{table}
\caption{Examples of stage 1 \& 2 reactions.}
\label{tab:reactions}
\begin{tabular}{lllll}
\hline
 \multicolumn{3}{c}{Reactants} & Product & Name \\
\cline{1-3}
\hline
OH & CH$_3$ & H$_2$O & CH$_3$OH & methanol \\
CH$_3$ & CHO & H$_2$O & CH$_3$CHO & acetaldehyde \\
CH$_2$ & NH & H$_2$O & CH$_2$NH & methylenimine \\
NH$_2$ & CHO & H$_2$O & NH$_2$CHO & formamide \\
CHO & CH$_3$O & H$_2$O & CH$_3$OCHO & methyl formate \\
\hline
OH & CH$_2$CH & H$_2$O & CH$_2$CHOH & vinylalcohol \\
CH$_3$ & CH$_2$NH$_2$ & H$_2$O & CH$_3$CH$_2$NH$_2$ & ethylamine \\
CH$_2$ & CHNH$_2$ & H$_2$O & CH$_2$CHNH$_2$ & vinylamine \\
NH$_2$ & N$_2$H$_3$ & H$_2$O & NH$_2$NHNH$_2$ & triazine \\
CHO & CH$_2$CH & H$_2$O & CH$_2$CHCHO & acrolein \\
\hline
\end{tabular}
\end{table}

Although hypothetical, there 
is some laboratory evidence to justify that the proposed reaction scheme is 
viable \citep[e.g.,][]{ND07}.
The products are almost entirely familiar, and many
are detected species in hot and warm cores and in the Central Molecular Zone of
the Milky Way \citep{Jetal12}. In particular, propylene should form by
the association of CH$_2$ and CH$_3$ (respectively) with the two first-stage
products listed above: CH$_3$CH and CH$_2$CH.

\section{Computations and results}

As stated above, at the densities that we are considering, the chemistry is
completely dominated by three-body reactions.
We therefore do not include any two-body, photochemical or cosmic-ray induced
reactions in our chemical network.
Also, as in our previous studies, we note that there are no clear observations 
of sulfur-bearing species in ice mantles and so we do not include any complex 
sulfur-bearing species.

We also follow the practice of \citet{RWV13} and adopt a single value for 
the rate coefficient (k$_{3B}$) for all reactions. This incorporates any 
implicit dependence on the temperature. 
Our previous work \citep{RWV13} has shown that the temperature-dependence 
of the rate coefficients is relatively unimportant, although at high 
temperatures the chemistry would tend to yield a thermochemical equilibrium 
composition.
An inspection of databases for known three-body reactions \citep[e.g.,][]{W07}
reveals that for reactions between radicals, the rate coefficients can be 
quite large ($> 10^{-26}$ cm$^6$s$^{-1}$), whereas for reactions involving
saturated species they may be significantly smaller ($\sim 10^{-33}-10^{-31}$ 
cm$^6$s$^{-1}$). Although we do not include the latter, we have considered 
values of k$_{3B}$ that cover the range $10^{-32}-10^{-26}$ cm$^6$s$^{-1}$.

We have developed two types of model to study the viability of propylene 
formation in high density sublimates: the first (1) calculates the 
time-dependence of the chemistry using one set of values for the free parameters. 
The second (2) calculates the final (asymptotic) abundance of one
selected species (typically propylene) for a large grid (typically 
$9\times 9\times 9\times 9$) of combinations of 
four parameters, spanning the ranges indicated in 
Table~\ref{tab:parameters}.

Examples of the final abundances relative to H$_2$O (for selected species)
obtained from the first model are given in Table~\ref{tab:model1}. In this calculation, the
initial density ($n_0$) is $1.0\times 10^{22}$cm$^{-3}$, the gas is allowed 
to expand freely ($\epsilon =1$), the `universal' rate coefficient ($k_{3B}$) 
is $1.0\times 10^{-29}$ cm$^3$s$^{-1}$ and the fraction of the mantle ice
converted into radicals ($F_{rad}$) is 1 percent.
The time-dependence of the abundances is straightforward;- a rapid rise is 
followed by levelling off to asymptotic values defined by chemical saturation
and/or geometrical dilution.
The species whose abundances are given in Table~\ref{tab:model1} have all been 
detected in the ISM. The abundances are all quite large,
showing that effective conversion of carbon to COMs is taking place.
However, this is for one single explosion event and cannot be 
directly compared to values determined from observations.
To do that requires a knowledge of the frequency of the explosions and how the
chemistry evolves in the ISM in the inter-explosion periods. In most cases, the
gas-phase chemistry of these species is very poorly understood.

\begin{table}
\begin{center}
\caption{Final abundances (relative to H$_2$O) for selected species obtained
from Model 1. The nomenclature $a(b)$ implies a value of $a\times 10^b$.}
\label{tab:model1}
\begin{tabular}{ll|ll}
\hline
Species  & Abundance & Species & Abundance \\ 
\hline
CH$_3$OH & 1.3(-4) & H$_2$CO & 9.3(-5) \\
HCOOH & 1.0(-4) & CH$_3$NH$_2$ & 2.4(-6) \\
CH$_3$CHO & 7.2(-6) & CH$_3$OCH$_3$ & 1.5(-6) \\
C$_2$H$_5$OH & 6.3(-6) & CH$_2$NH & 1.9(-6) \\
NH$_2$CHO & 1.9(-6) & CH$_3$OCHO & 1.2(-6) \\
CH$_2$OHCHO & 1.3(-5) &  (CH$_2$OH)$_2$ & 1.2(-5) \\
CH$_2$CHOH & 3.2(-6) & CH$_3$CH$_2$CHO & 9.5(-7) \\
CH$_3$CH$_2$CH$_2$OH & 7.8(-7) & {\bf CH$_3$CHCH$_2$} & {\bf 1.9(-6)} \\
CH$_3$CHNH & 7.6(-7) &  & \\
\hline
\end{tabular}
\end{center}
\end{table}

From Table~\ref{tab:parameters} we see that there
are several parameters in the model. However, the three-body chemical 
timescale is $t_{chem}\sim 1/(k_{3B}n_0^2)$. 
From our calculations we indeed find that the results depend on the 
value of $k_{3B}n_0^2$ and not independently on $k_{3B}$ and $n_0^2$,
so that the product $k_{3B}n_0^2$ is a single free parameter. 
The values of $k_{3B}$ and $n_0$ used to obtain the results given in
Table~\ref{tab:model1} imply a value of $k_{3B}n_0^2 = 10^{15}$s$^{-1}$.
The free parameters are thus; the ice composition, the branching ratios for
radical production, $F_{rad}$, $k_{3B}n_0^2$ and $\epsilon$.

Fixing the chemical parameters as per Table~\ref{tab:parameters},
we present in Table~\ref{tab:model2a} the final 
(asymptotic) abundances for propylene as a function of two free parameters; 
the fraction of the ice that is converted to radicals ($F_{rad}$) and 
$k_{3B}n_0^2$, for $\epsilon =1$.

\begin{table}
\begin{center}
\caption{Results obtained from model 2 with $\epsilon=1.0$:
log of the final abundances (relative to H$_2$O) for CH$_3$CHCH$_2$.}
\label{tab:model2a}
\begin{tabular}{llllll}
\hline
 & \multicolumn{5}{c}{Log$_{10}$ F$_{rad}$} \\
\cline{2-6} 
 Log$_{10}$($k_{3B}n_0^2$) & -4.00 & -3.50 & -3.00 & -2.50 & -2.00 \\
\hline
10.0 & -16.86 & -15.36 & -13.86 & -12.37 & -10.87 \\
11.0 & -14.86 & -13.37 & -11.87 & -10.39 & -8.94 \\
12.0 & -12.87 & -11.39 & -9.94 & -8.57 & -7.34 \\
13.0 & -10.94 & -9.57 & -8.34 & -7.29 & -6.39 \\
14.0 & -9.34 & -8.29 & -7.39 & -6.60 & -5.90 \\
15.0 & -8.39 & -7.60 & -6.90 & -6.28 & -5.73 \\
16.0 & -7.90 & -7.28 & -6.73 & -6.22 & -5.72 \\
17.0 & -7.73 & -7.22 & -6.72 & -6.22 & -5.72 \\
18.0 & -7.72 & -7.22 & -6.72 & -6.22 & -5.72 \\
\hline
\end{tabular}
\end{center}
\end{table}

Table~\ref{tab:model2b} shows the same as Table~\ref{tab:model2a}, but for
the case of significant gas trapping/inhibited expansion ($\epsilon=0.1$).
This effectively allows the gas to evolve for longer at the highest densities.
However, from equation~\ref{eqn:dynamics} we can see that the dynamical
(geometrical dilution) timescale is of the order of 
$t_{dyn}\sim(1/\epsilon)$\,ns. 
The efficiency of the conversion of the radicals trapped in the ice to 
complex organic molecules is effectively determined by the ratio of this
timescale to the chemical timescale.
A saturation limit will apply in those situations where $t_{chem}<<t_{dyn}$.
Thus, the differences between these two tables are only significant for 
the smaller values of $k_{3B}n_0^2$ where the above inequality does not hold.

\begin{table}
\begin{center}
\caption{Results obtained from model 2 with $\epsilon=0.1$:
log of the final abundances (relative to H$_2$O) for CH$_3$CHCH$_2$.}
\label{tab:model2b}
\begin{tabular}{llllll}
\hline
 & \multicolumn{5}{c}{Log$_{10}$ F$_{rad}$} \\
\cline{2-6} 
 Log$_{10}$($k_{3B}n_0^2$) & -4.00 & -3.50 & -3.00 & -2.50 & -2.00 \\
\hline
10.0 & -14.89 & -13.39 & -11.90 & -10.41 & -8.96 \\
11.0 & -12.90 & -11.41 & -9.96  & -8.59  & -7.36 \\
12.0 & -10.96 & -9.59  & -8.36  & -7.30  & -6.40 \\
13.0 & -9.36  & -8.30  & -7.40  & -6.61  & -5.91 \\
14.0 & -8.40  & -7.61  & -6.91  & -6.28  & -5.73 \\
15.0 & -7.91  & -7.28  & -6.73  & -6.22  & -5.72 \\
16.0 & -7.73  & -7.22  & -6.72  & -6.22  & -5.72 \\
17.0 & -7.72  & -7.22  & -6.72  & -6.22  & -5.72 \\
18.0 & -7.72  & -7.22  & -6.72  & -6.22  & -5.72 \\
\hline
\end{tabular}
\end{center}
\end{table}

Some clear trends can be seen in the results from these models. Firstly, from
Tables~\ref{tab:model2a} and \ref{tab:model2b}, the saturation effect referred
to above is discernable; a robust limiting value for the propylene abundance
(relative to H$_2$O) of $Y_{sat.}\sim 1.9\times 10^{-6}$ is obtained. 
We find that this saturation limit is primarily dependent on the chemical 
initial conditions, and is independent of the physical parameters; 
$n_0$, $k_{3B}$ and $\epsilon$. Thus,
\begin{equation}
Y_{sat.} \sim 1.9\times 10^{-6} \left(\frac{F_{rad}}{1\%}\right)
\left(\frac{x(CH_4)}{4\%}\right)^{2.2}
\left(\frac{f_{OH}}{50\%}\right)^{-0.5}
\label{eqn:saturate}
\end{equation}
where $x(CH_4)$ is the fractional abundance of CH$_4$, relative to H$_2$O in 
the ice and $f_{OH}$ is the fraction of the radicals produced from the 
photodissociation of H$_2$O that are in the form of OH. 
This saturation limit, corresponding to the complete conversion of radicals to
molecular species, is set by a number of factors, including;
\begin{enumerate}
\item The fact that there are a large number of species in the network, 
none of which is formed preferentially (due to the use of a single value of 
$k_{3B}$ for all reactions). The carbon has to be shared between these species.
\item The (relatively small) molecules that are formed in the first stage 
do not react further and also act a sink for carbon.
\item OH is the most abundant radical. Many of the reactions leading to 
hydrocarbon growth will be in competition with reactions involving OH. 
\end{enumerate}
In the general situation, we can see (from Tables~\ref{tab:model2a} and 
\ref{tab:model2b}) that the propylene abundance (Y$_{prop.}$) is approximately
within an order of magnitude of $Y_{sat.}$ when $F_{rad}>0.001$ and 
$k_{3B}n_0^2 > 10^{14}$\,s$^{-1}$.

\section{Discussion and conclusions}

The explosion chemistry is completed on a nanosecond timescale. The 
products of this chemistry may not be directly observed, but are injected 
into the dense gas of a molecular cloud core. Observational effects are 
therefore not necessarily defined by the explosion chemistry, but are 
determined by the injection and subsequent chemistry in the dark cloud. Using 
the result from the previous section, it now becomes possible to estimate an 
injection rate for propylene into a dense cloud core like those in TMC-1.
Comparing this timescale to that for the destruction of propylene by gas-phase
chemistry in the molecular cloud, we can use a simple argument to estimate the
time-averaged abundance of propylene that could result from this mechanism.

In a dark, quiescent, environment nearly all heavy species accrete onto dust
grains, on a timescale of the freeze-out time;
$t_{fo} \sim (3\times 10^{16}/{n_H})$s,
where $n_{H}$ is the total (hydrogen nucleon) density \citep{RHMW92}. 
In the absence of an effective desorption mechanism (such as explosions), 
molecular material would be mostly depleted from the gas phase, so it is 
fair to assume - in the context of our model - that the ice mantle explosion
occurs soon after time $t_{fo}$ has elapsed; effectively limiting the ice 
mantle growth.

A large fraction of the ice mantle is composed of H$_2$O.
Thus, the average number density of molecules ejected in one 
freeze-out/ejection cycle is $\sim X_{H_2O}.n_H$, where $X_{H_2O}$ is the 
fractional abundance of H$_2$O (relative to $n_H$).
The mean injection rate of propylene molecules is therefore $R_{prop.}$, given
by
\[ R_{prop.} \sim X_{H_2O}.\frac{n_{H}.Y_{prop.}}{t_{fo}} 
{\rm ~~cm^{-3}s^{-1}} \]
where $Y_{prop.}$ is the fractional abundance of propylene (relative to H$_2$O)
that is formed in the high density explosion chemistry.
Thus,
\[ R_{prop.} \sim \frac{X_{H_2O}n_{H}^2 Y_{prop.}}{3\times 10^{16}} 
{\rm ~~cm^{-3}s^{-1}.} \]
The mean number density of propylene molecules is then given by
$ n_{prop.} = R_{prop.}.t_{loss} $,
where $t_{loss}$ is the typical lifetime of propylene in the gas phase.
We can estimate the destruction timescale for propylene by considering the main
reaction channels in interstellar dark cloud conditions. 
From a simple model of the quiescent phase chemistry of TMC-1 
which includes gas-grain interactions (freeze-out and non-thermal
desorption) we find that the principal destruction channels are:-
\[ {\rm CH_3CHCH_2 + H_3^+ \to products} \]
\[ {\rm CH_3CHCH_2 + C \to C_4H_3 + H_2 + H} \]  
\[ {\rm CH_3CHCH_2 + CN \to products} \]
\[ {\rm CH_3CHCH_2 + O \to products} \]
The destruction timescale for propylene by any particular channel is of the 
order of $\sim 1/(k.n_r)$, where $k$ is the rate coefficient and $n_r$ is the
density of the reactant. Using data for these reactions from the UMIST 
Database For Astrochemistry \citep{udfa12}, 
\citep[and][for the last reaction]{Sab07} 
and the predicted abundances from our model, we find that the 
destruction timescale is typically $t_{loss}\sim 10^4-10^5$years.

The mean fractional abundance of propylene (relative to hydrogen) is
\[ X_{prop.} = \frac{n_{prop.}}{n_H} = 
\frac{X_{H_2O} n_{H} Y_{prop.} t_{loss}}{3\times 10^{16}} \]
Using values of $X_{H_2O}=10^{-4}$, $n_{H}=10^5$cm$^{-3}$, 
$Y_{prop.}=2\times 10^{-6}$ and $t_{loss}=10^4-10^5$years,
we obtain $X_{prop.}\sim 2\times 10^{-10} - 2\times 10^{-9}$.
This is comparable to the observationally inferred value in TMC-1,
although it requires the explosion and re-cycling mechanism to be 
operating at near-optimal efficiency.

This discussion shows that rather special conditions may be 
required to produce $X_{prop.}$ of this magnitude. 
The efficiency of propylene formation (in the saturation limit) is essentially 
independent of the physical parameters in the model and is defined by the ice
mantle composition (and the degree of processing to radicals).
The actual, observed, abundances in the dark cloud phase are determined by this 
value and the ratio of the explosion to gas-phase destruction timescales.
We may now speculate as to why propylene is detected in TMC-1, but not in 
other sources, such as Orion-KL.
From the above arguments, one or more of the following possibilities may be
valid:
\begin{enumerate}
\item TMC-1 may be chemically young \citep{HWV01} in which case one would 
expect there to be more atomic carbon in the gas phase. If so, then 
subsequent to the hydrogenation of accreted carbon to CH$_4$ the ice
mantles may have a higher CH$_4:$H$_2$O ratio than normal,
\item The dust grain properties may affect the ice composition. Thus, if the 
ices were warmer than normal 
then, on the basis of the relative adsorption energies,
the ratio of CH$_4$ to CO may be increased. 
This could, in effect, increase the C$:$O ratio in the explosion
chemistry and enhance the formation of hydrocarbons, 
\item In the high mass star-forming region of Orion the chemistry may
be dominated by the presence of shocks. This could result in higher 
abundances of atomic oxygen and therefore shorter loss timescales for
propylene.
\end{enumerate}

There are obviously a large number of approximations in our model, most 
significantly the reaction network is very speculative and the adoption of a
single value for the three-body reaction rate coefficients is a major
simplification. In reality there could be large differences between the 
rate coefficients which would result in strong variations in the formation
efficiencies and abundance ratios.
However, our model is based on plausible assumptions and we can conclude that:-
\begin{enumerate}
\item Given that, under normal interstellar conditions, a gas-phase chemistry
is unable to explain the observed abundances of COMs, an alternative mechanism
for their formation is required that (a) is highly efficient at converting
carbon to COMs, {\em and} (b) ensures that the COMs so-produced are efficiently
transmitted to the (ambient) gas-phase.
\item A chemistry rich in COMs may arise from the explosion of ice mantles.
\item Provided 0.1 percent or more of the ice is converted to radicals and
the product of the reaction rate coefficients with the square of the initial,
post-sublimation, gas density is $>10^{14}$\,s$^{-1}$, the abundances of the COMs
may approach saturation levels. 
This criterion is physically plausible, and does not imply that the process 
would only be efficient if we adopt extreme values of the free parameters.
\item The saturation levels are defined by the ice composition, as specified in
equation~\ref{eqn:saturate}, and not the physical characteristics of the 
mantle explosion.
\item The injection rate may be sufficiently high to affect abundances in the
cold cloud environment.
\item However, the injection rate is sensitive to various parameters that
may vary significantly in different regions.
\item Other scenarios in which association chemistry may be important almost
certainly exist. For example, radical association chemistry may be 
important in regions containing recently evaporated ices, such as 
so-called hot cores. These regions have molecular hydrogen number 
densities $\sim 10^7$ cm$^{-3}$, and stabilisation of the complexes will be 
radiative rather than collisional.
\end{enumerate}

\bibliographystyle{mn2e}

\end{document}